\documentclass[pdflatex,sn-mathphys-num]{sn-jnl}

\usepackage{graphicx}%
\usepackage{multirow}%
\usepackage{amsmath,amssymb,amsfonts}%
\usepackage{amsthm}%
\usepackage{mathrsfs}%
\usepackage[title]{appendix}%
\usepackage{xcolor}%
\usepackage{textcomp}%
\usepackage{manyfoot}%
\usepackage{booktabs}%
\usepackage{algorithm}%
\usepackage{algorithmicx}%
\usepackage{algpseudocode}%
\usepackage{listings}%
\usepackage{subcaption} %
\usepackage{float}
\usepackage{tabularx} 
\usepackage[section]{placeins}
\usepackage{tabularray}
\usepackage[utf8]{inputenc}
\usepackage[T1]{fontenc}
\usepackage{epstopdf}
\usepackage{caption} 
\UseTblrLibrary{booktabs}
\usepackage{geometry}
\theoremstyle{thmstyleone}%
\theoremstyle{thmstyletwo}%

\theoremstyle{thmstylethree}%

\begin{document}

\title[Article Title]{Estimation of the Inverse Compton Scattering Background in MeV Gamma-Gamma Collider}


\author[1]{\fnm{Ping} \sur{Zhou}}\email{zhoup73@mail2.sysu.edu.cn}

\author*[1]{\fnm{Mei-Yu} \sur{Si}}\email{simy@mail.sysu.edu.cn}

\author*[1]{\fnm{Yuan-Jie} \sur{Bi}}\email{biyj@mail.sysu.edu.cn}

\author[2]{\fnm{Illya} \sur{Drebot}}\email{illya.drebot@mi.infn.it}

\author[1]{\fnm{Yong-Sheng} \sur{Huang}}\email{huangysh59@mail.sysu.edu.cn}

\affil[1]{\orgdiv{School of Science}, \orgname{Shenzhen Campus of Sun Yat-sen University}, \orgaddress{\street{Gongchang Road 66}, \city{Shenzhen}, \postcode{518107}, \state{Guangdong Province}, \country{China}}}

\affil[2]{\orgname{INFN-Milan (Istituto Nazionale di Fisica Nucleare, Sezione di Milano)}, \orgaddress{\street{Via Celoria 16}, \city{Milan}, \postcode{20133}, \state{Milan}, \country{Italy}}}


\abstract{The MeV Gamma-Gamma Collider would provide a direct experimental platform for elastic light-by-light scattering ($\gamma\gamma \to \gamma\gamma$) and the Breit-Wheeler process with two real photons ($\gamma\gamma \to e^+e^-$). A Monte Carlo code, the Genie Background Evaluation Tool (GBET), has been developed to fully simulate two successive inverse Compton scatterings in the linear regime, including $e^- + \text{laser} \to e^-+ \gamma$ and \( e^- + \gamma \rightarrow e^- + \gamma \). GBET overcomes the inherent information loss in traditional luminosity-spectrum-based chain simulations, preserving full particle-level information and achieving higher physical fidelity. The effectiveness of the code is verified by benchmarking against the simulation results of CAIN. GBET shows that the event rate of background photons generated by the first inverse Compton scattering is \( \textnormal{6.24} \times \textnormal{10}^{\textnormal{7}} \)/s, with energies below 18 eV; the second inverse Compton scattering generates background electrons at 51.99/s and photons at 0.99/s, both with energies below 11 MeV. In addition, Møller scattering contributes background electrons at 0.56/s with energies around 200 MeV. The count rates of background electrons and positrons originating from the Breit-Wheeler process are 1312.2/s and 1314.3/s, respectively, with energy distributions ranging from 511 to 720 keV.}




\maketitle
\section{Introduction}\label{sec1}

Gamma-Gamma Collider is a novel experimental device that provides a particularly clean environment to experiments in high-energy physics, offering unique and rich physical opportunities at different energy scale. It has not yet been realized, but the concept has a long and rich history spanning several decades. Its basic principle is to use high-energy electron beams and strong laser pulses to generate gamma-ray beams through inverse Compton scattering \cite{1,2,3,4}. Subsequently, two gamma-ray beams were used to collide to study new physical phenomena. In the 1980s, the Novosibirsk Institute of Nuclear Physics proposed the idea of Gamma-Gamma Collider based on the VLEPP linear collider \cite{5}. Later, it was considered as an ``add-on'' option to high-energy electron-positron linear colliders, such as SLC, JLC, NLC, TESLA and ILC \cite{6,7,8}. All the proposals above share a common feature: exploring the energy frontier, at the scale of a Higgs factory or beyond. The realization of Gamma-Gamma Collider at such high energies depends on high-energy electron accelerators and high-collimation, high-power short-pulse laser technology. The associated experimental facilities and technologies require considerable time to be further developed and refined.

In view of this, scientists have discussed various Gamma-Gamma Collider plans and proposed the idea of building a low-energy Gamma-Gamma Collider at the 2017 Future Photon Collider Workshop and the ICFA miniworkshop \cite{9,10}. W. Chou et al. proposed the construction of a Gamma-Gamma Collider with a center-of-mass (CM) energy of 1-2 MeV. This energy region lies close to the peak of elastic light-by-light scattering cross section ($\gamma\gamma \to \gamma\gamma$) \cite{11} and just above the threshold for Breit-Wheeler pair production ($\gamma\gamma \to e^+e^-$) \cite{12}. Both processes were predicted in the 1930s but have not yet been directly observed or measured, apart from the experiment of Burke et al., which detected the more complex multiphoton Breit-Wheeler process over twenty years ago \cite{13}. In this regime, the cross section for elastic light-by-light scattering can reach the order of $\mu$b \cite{14,15}, while the Breit-Wheeler pair production cross section is on the order of 100 $\mathrm{mb}$. It offers a promising opportunity for direct experimental tests of these two QED processes in the laboratory. T. Takahashi et al. studied the feasibility of experimental verification, showing that elastic light-by-light scattering can be observed with a statistical significance of $5\sigma$ within half a year of data collection \cite{16}. Through start-to-end simulations, Drebot et al. demonstrated that the Breit-Wheeler pair production process is detectable \cite{17}.

As distinguishing signal from background is a key experimental challenge, precise background analysis is crucial. The background can be categorized into three categories: (1) the natural cosmic ray background, which can be suppressed through triggering; (2) backgrounds from other physical processes at the interaction point (IP); (3) the beam scattering background. Specifically, the latter two categories include: triplet pair production ($\gamma e^- \to e^+e^-e^-$) \cite{21}, second inverse Compton scattering ($e^-\gamma \to e^-\gamma$) and Møller scattering ($ e^-e^- \to e^-e^-$) \cite{22}. These processes occur simultaneously with elastic light-by-light scattering and Breit-Wheeler production at the IP in Gamma-Gamma Collider. The particles produced by these processes may also be captured by the detector, thus forming background signal. To date, research on background analysis is still relatively scarce \cite{16,17,18,19,20}. Moreover, the conventional approach relies on CAIN's~\cite{29} luminosity spectrum as input, which inherently lacks full particle-level information. To overcome this limitation, a Monte Carlo (MC) code, GBET, is developed to study the latter two backgrounds. The validity is confirmed through benchmark comparisons with both the theoretical integrals and the simulation results from CAIN. This study focuses on the background estimation caused by the two inverse Compton scattering processes occurring in the MeV Gamma-Gamma Collider, including the first inverse Compton scattering between an electron and a laser photon ($e^- + \text{laser} \to e^-+ \gamma$), and the second inverse Compton scattering between an electron and a gamma photon ($e^- + \gamma \to e^- + \gamma$). Based on the parameters of the MeV Gamma-Gamma Collider, two inverse Compton scattering processes in the linear regime were simulated, providing the event rates and energies of background particles. This study can provide a reference for detectors to distinguish between signals and background and promote the construction of Gamma-Gamma Collider.

The structure of this paper is as follows. Section \ref{sec2} presents an overview of the MeV Gamma-Gamma Collider, outlines the beam parameters, and introduces the theoretical foundation --- inverse Compton scattering. Section \ref{sec3} focuses on the number and energy of background particles produced by two successive inverse Compton scattering, and briefly discusses background particles from other processes. Section \ref{sec4} introduces the MC code in detail and provides benchmark comparisons. Finally, Section \ref{sec5} summarizes our work.

\section{MeV Gamma-Gamma Collider}\label{sec2}

The Gamma-Gamma Collider is a new type of high-energy physics experimental device, which has unique and rich physical targets in all energy regions. It complements the Hadron Collider, Electron-Positron Collider, etc. and plays an irreplaceable role. The MeV Gamma-Gamma Collider aims to achieve three major scientific objectives: (1) the realization of the first Gamma-Gamma Collider; (2) the first observation and measurement of real photon-photon collisions and scattering in the laboratory; (3) the first observation and measurement of the Breit-Wheeler process involving two real photons in the laboratory. The platform also holds potential for searches of axions and dark photons in the MeV range \cite{39,40}. Moreover, it can be expanded to develop six application platforms supporting research and development in high energy density physics and ultrafast imaging, demonstrating both scientific significance and application potential.

\subsection{Collider parameters}\label{subsec:section2.1}

The realization of Gamma-Gamma Collider with CM energy of 1-2 MeV is based on the following two steps: first, electrons and lasers produce high-energy gamma-ray through inverse Compton scattering; then the high-energy gamma-ray beams on both sides collide at IP to study new physical phenomena.   

\begin{figure}[H]
	\centering
	\includegraphics[width=1\textwidth]{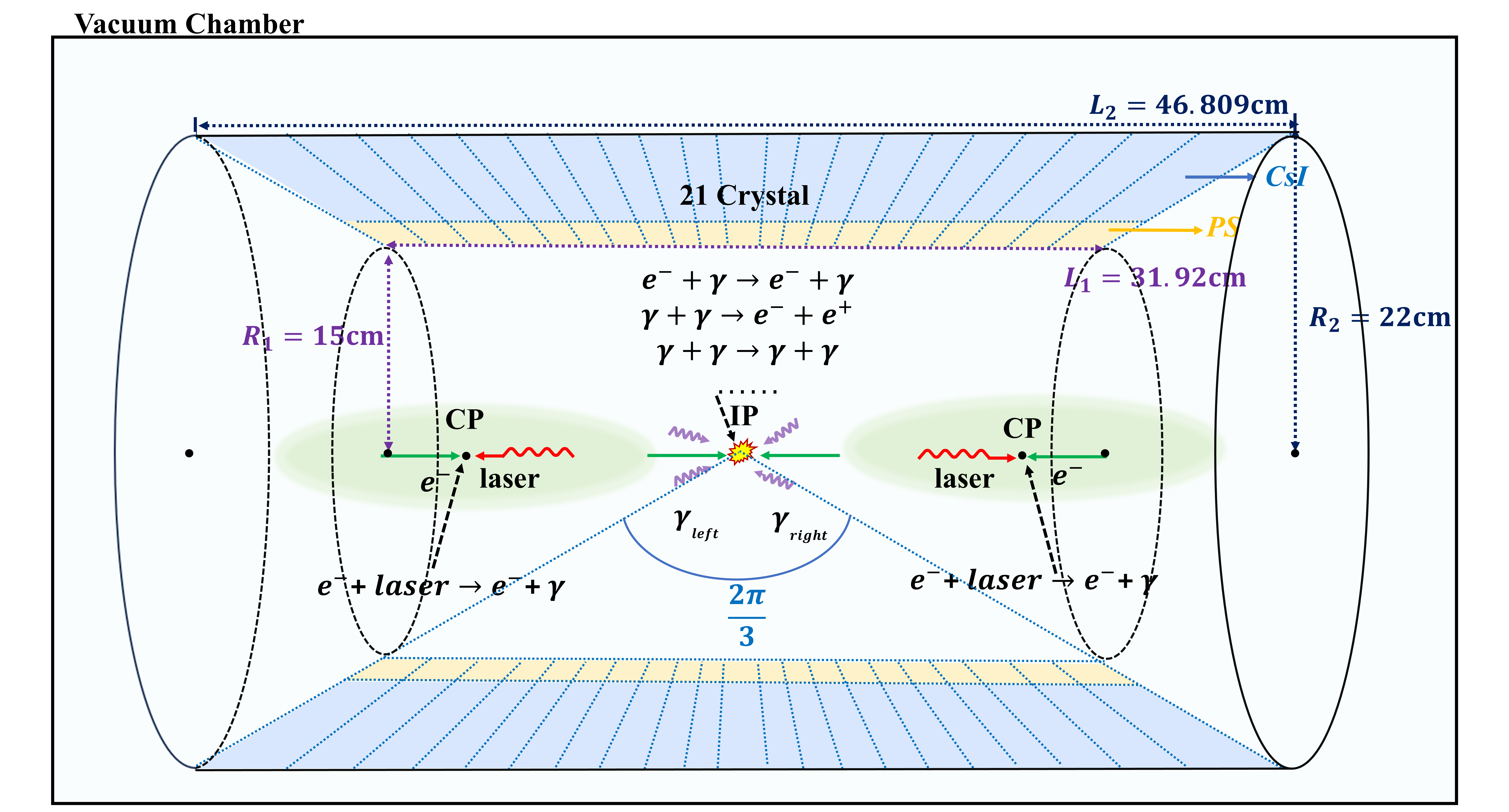}
	\caption{Schematic diagram of colliding beams in the Gamma-Gamma Collider. The laser pulse and electron beam are scattered at the CP to produce Compton photons. The Compton photons and electrons are brought to the IP and produce $e^-\gamma$, $\gamma\gamma$ and $e^-e^-$ collisions. The detector surrounding the beams consists of a scintillator array composed of CsI and plastic scintillators (PS), with inner and outer radii of \( R_1 = 15 \, \text{cm} \) and \( R_2 = 22 \, \text{cm} \), respectively, and lengths of \( L_1 = 31.92 \, \text{cm} \) and \( L_2 = 46.809 \, \text{cm} \). The entire detector is located inside a vacuum chamber. The angular range of the detector is set to $\pi / 6$ to $5\pi / 6$, particles in this angular range will be detected.
	}\label{Fig1}
\end{figure}

The principle of Gamma-Gamma Collider, as illustrated in Fig.~\ref{Fig1}, consists of two conversion points (CP) and one IP. The distance from CP to IP is 383 $\mu\mathrm{m}$. Two electron beams pass through the final focusing system and are focused to 2 $\mu\mathrm{m}$ in the transverse dimension at the IP. At the CP, the electron beams collide head-on with the laser beams, generating high-energy gamma-ray beams (several hundred keV) through inverse Compton scattering. Then the gamma-ray beams from both sides collide at IP, resulting in elastic light-by-light scattering, which occurs simultaneously with the other processes mentioned in Section~\ref{sec1}. The angular range of the detector is set to $\pi / 6$ to $5\pi / 6$, particles in this angular range will be detected. Table~\ref{tab1} lists the laser and electron parameters used in the MeV Gamma-Gamma Collider. This project uses 200 MeV, 2 nC electron beams with 600 $\mu\mathrm{m}$ bunch length, 2 $\mu\mathrm{m}$ transverse size, and 50 Hz repetition rate. The lasers have a 1054 nm wavelength, 2 J pulse energy, 10 $\mu\mathrm{m}$ waist, 1 ps pulse length, and operates at 50 Hz. Based on the parameters in Table~\ref{tab1}, GBET shows that the total yield of Compton photons is about $1.48 \times 10^{12}$/s; The maximum CM energy of $\gamma$$\gamma$ collision is $\sqrt{s_{\gamma\gamma}} \approx 1.44~\mathrm{MeV}$; The $\gamma\gamma$ collision luminosity $\mathcal{L}_{\gamma\gamma}$ achieves $5.10 \times 10^{28}\;\mathrm{cm^{-2}s^{-1}}$,
which exceeds both the $e^{-}e^{-}$ luminosity ($1.18 \times 10^{28}\;\mathrm{cm^{-2}s^{-1}}$)
and the $e^-\gamma$ luminosity ($2.38 \times 10^{28}\;\mathrm{cm^{-2}s^{-1}}$). 
\renewcommand{\arraystretch}{1.25}
\begin{table}[htbp!]
\centering
\caption{Laser and Electron parameters for the MeV Gamma-Gamma Collider. The laser pulse's size and length were defined as the Root Mean Square (RMS) of the intensity \cite{23}}
\label{tab1}

\noindent
\begin{tabularx}{\textwidth}{
    >{\centering\arraybackslash\hsize=1.5\hsize}X
    >{\centering\arraybackslash\hsize=0.5\hsize}X
    >{\centering\arraybackslash\hsize=1.5\hsize}X
    >{\centering\arraybackslash\hsize=0.5\hsize}X
}
\toprule
Laser Parameters & Values & Electron Parameters & Values \\
\midrule
Wave Length ($\mu$m)           & 1.054          & Energy (MeV)               & 200            \\ 
Size at Focus ($\mu$m)         & 5              & Bunch Charge (nC)          & 2              \\ 
Rayleigh Range ($\mu$m)        & 298            & Beam Spot Size (IP) ($\mu$m)    & 2              \\ 
Pulse Energy (J)               & 2              & Bunch length ($\mu$m)      & 600            \\ 
Pulse Length (ps)              & 1              & Beta at IP ($\mu$m)              & 625            \\ 
Repetition (Hz)                & 50             & Emittance (nm)             & 6.4            \\ 
Angle to e-beam (mr)           & 0              & Crossing angle (mr)         & 0              \\ 
IP-CP distance ($\mu$m)        & 383            & Repetition (Hz)            & 50             \\ 
\bottomrule
\end{tabularx}

\end{table}

\subsection{Inverse Compton scattering}\label{subsec:section2.2}

Compton scattering usually refers to the process of elastic scattering of photons with stationary electrons, while the process of low-energy photons interacting with high-energy electrons and gaining energy is called inverse Compton scattering or Compton backscattering \cite{24}. In Gamma-Gamma Collider, two types of inverse Compton scattering processes are involved. The first occurs in the Thomson regime, where the electrons and the laser photons interact. The second takes place in the Compton regime, involving electrons interacting with high-energy gamma photons. The two show significant differences in energy transfer characteristics and total cross section.
\subsubsection{Scattered photon energy}\label{subsubsec:section2.2.1}
When an electron $ p = (E_e / c, \vec{p}) $ collides with a photon $ k = (E_L / c, \vec{k}) $ at an arbitrary angle, according to the conservation of four-momentum:

\begin{equation}
p + k = p' + k',
\label{eq:1} 
\end{equation}
where \(p\) and \(k\) are the four-momenta of the electron and photon respectively before the interaction, and \(p'\) and \(k'\) are the four-momenta after the interaction.
The scattered photon energy $E_L'$ is given by
\begin{equation}
E_L' = E_L \cdot \frac{1 - \beta \cdot \cos{\theta_i}}{1 - \beta \cdot \cos{\theta_f} + (E_L / E_e) \cdot (1 - \cos{\theta_{ph}})},
\label{eq:2} 
\end{equation}
where \(\theta_i\) denotes the angle between the electron momentum \(\vec{p}\) and the initial photon direction \(\vec{k}\), \(\theta_f\) represents the angle between the electron momentum \(\vec{p}\) and the scattered photon direction \(\vec{k}'\), and \(\theta_{ph}\) is the angle between the photon directions before and after scattering.
For a head-to-head collision, \(\theta_i = \pi\) and \(\theta = \theta_f\ = \pi - \theta_{ph}\), the energy of the scattered photon is only related to \( \theta \), and can be expressed as

\begin{equation}
E_L' = \frac{(1 + \beta)E_L}{(1 - \beta \cos \theta) + (1 + \cos \theta)E_L / E_e}.
\label{eq:3} 
\end{equation}
For an ultra-relativistic electron ($\gamma = E_e / (m c^2) \gg 1$) and a small scattering angle \((\theta \ll 1)\), Eq.~\eqref{eq:3} can be simplified to

\begin{equation}
E_L' \approx \frac{4\gamma^2 E_L}{1 + \gamma^2 \theta^2 + 4\gamma^2 E_L / E_e},
\label{eq:4} 
\end{equation}
More explicitly, the energy of the scattered photon reaches maximum value for \(\theta = 0\)

\begin{equation}
E_L'^{\text{max}} = \frac{4\gamma^2 E_L}{1 + 4\gamma^2 E_L / E_e}.
\label{eq:5} 
\end{equation}
Therefore, it can be obtained from Eq.~\eqref{eq:5} that the maximum energy of Compton photons produced by the head-on collision of 200 MeV electrons ($\gamma \approx 391$) and laser with a wavelength of 1054 nm ($E_L \approx 1.18\,\text{eV}$
) can reach 720 keV.

We can introduce the electron recoil factor in order to better understand the behavior of the scattering process in different energy regions, which is defined as:

\begin{equation}
R = \frac{4E_LE_e}{(mc^2)^2} = \frac{4\gamma E_L}{mc^2} = \frac{4\gamma^2 E_L}{E_e}
\label{eq:6} 
\end{equation}
The electron recoil factor $R$ is related to the energy available in the CM system $\sqrt{s_{e\gamma}}$
, through the expression $\sqrt{s_{e\gamma}} = mc^2 (1 + R)$. As can be seen from Eq.~\eqref{eq:5} and ~\eqref{eq:6}, it limits the maximum energy of scattered photon. From Eq.~\eqref{eq:5}, it can be seen that when \( R \ll 1 \), the maximum energy of the scattered photon is approximately \( 4\gamma^2 E_L \). In contrast, when \( R \gg 1 \), recoil is essential, and the maximum energy of the scattered photon approaches that of the incident electron. This energy is constrained by a kinematically determined upper limit, approximately \( E_e \) \cite{25}.

\subsubsection{Scattering cross section}\label{subsubsec:section2.2.2}

\begin{figure}[H]
\centering
\includegraphics[width=0.5\textwidth]{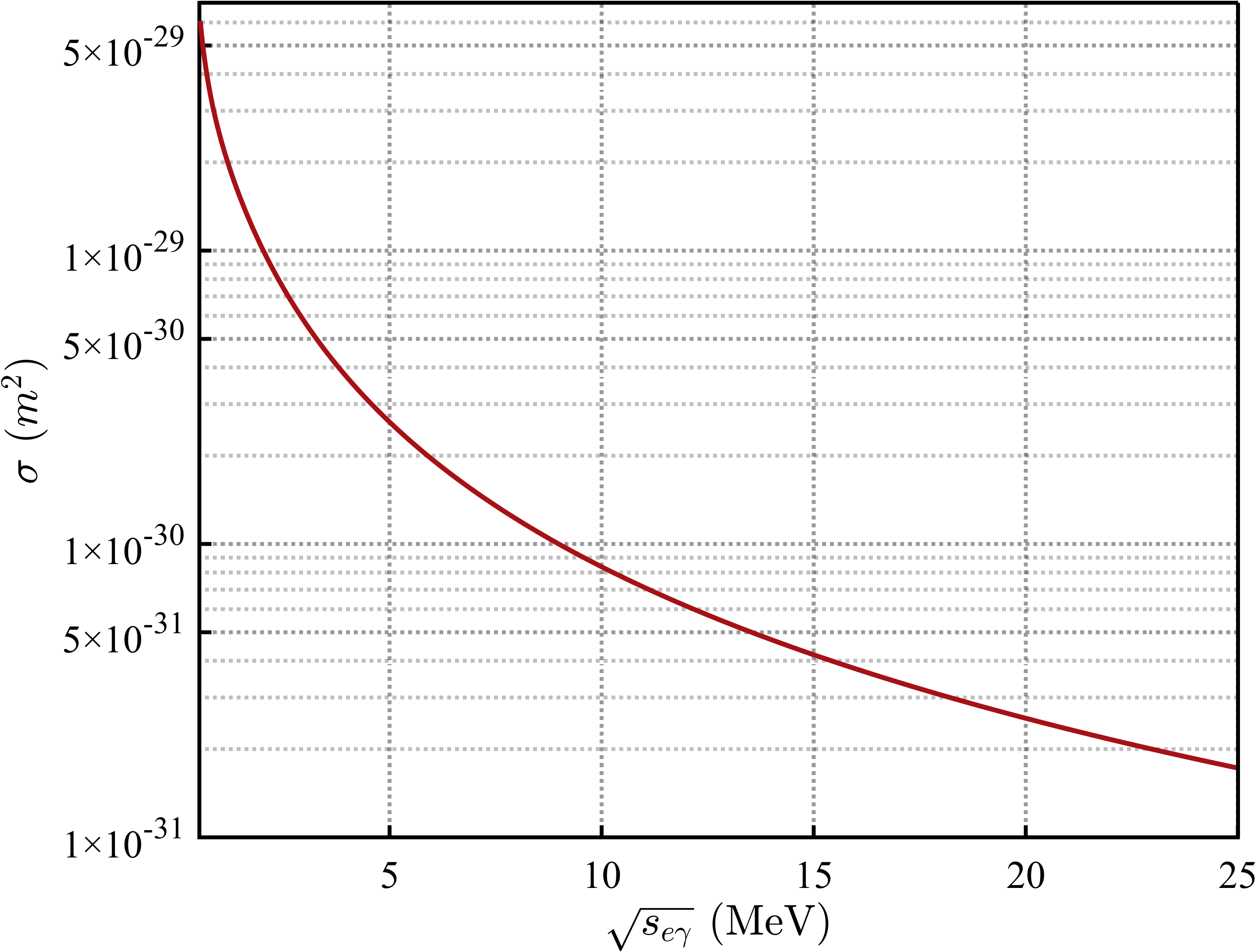} 

\caption{Cross section of inverse Compton scattering. Variation trend of total cross section with CM energy $\sqrt{s_{e\gamma}}$
: The CM energy for the first inverse Compton scattering is around 0.511 MeV, with a total cross section of $6.628 \times 10^{-29}$ $\mathrm{m^2}$. The CM energy for the second inverse Compton scattering ranges from 0.511 MeV to 24 MeV, with the corresponding total cross sections over a range from $6.628 \times 10^{-29}$ $\mathrm{m^2}$ to $1.855 \times 10^{-31}$ $\mathrm{m^2}$.}
\label{Fig2}
\end{figure}

The QED book \cite{26} defines the differential cross section of Compton scattering in terms of the Lorentz invariant form, where the differential cross section for unpolarized electrons and photons is given by
 
\begin{equation}
\frac{d\sigma}{dy d\phi_{\text{CM}}} = \frac{4r_e^2}{x^2}  \left\{
\left( \frac{1}{x} - \frac{1}{y} \right)^2 
+ \left( \frac{1}{x} - \frac{1}{y} \right) 
+ \frac{1}{4} \left( \frac{x}{y} + \frac{y}{x} \right)
\right\},
\label{eq:7} 
\end{equation}
where $r_e$  is the classical radius of the
electron, $ \phi_{\text{CM}} $ is the azimuthal angle in the CM frame, and the variables \(x\) and \(y\) are defined as:
\begin{equation}
x = \frac{s - (mc^2)^2}{(mc^2)^2}, \quad y = \frac{(mc^2)^2 - u}{(mc^2)^2}
\label{eq:8} 
\end{equation}
where \(s\), \(u\) are the Mandelstam variables \cite{26}
\begin{equation}
s = (p + k)^2, \quad u = (p - k')^2 = -\frac{(s^2 - (mc^2)^2)}{2s} + \frac{(s - mc^2)^2}{2s} \cos \theta_{CM} + (mc^2)^2.
\label{eq:9} 
\end{equation}
$\theta_{CM}$ is the scattering angle in the CM system, and \(x\) , \(y\) satisfy the following relation \cite{26}

\begin{equation}
\frac{x}{x + 1} \leq y \leq x.
\label{eq:10} 
\end{equation}
Substituting Eq.~\eqref{eq:8} and ~\eqref{eq:9} into Eq.~\eqref{eq:7}, the angular differential cross section is

\begin{equation}
\frac{d\sigma}{d\Omega} = \frac{2 r_e^2 (mc^2)^2}{s} \left[ 
\frac{1}{4} \left( \frac{x}{y} + \frac{y}{x} \right) + 
\left( \frac{1}{x} - \frac{1}{y} \right) + 
\left( \frac{1}{x} - \frac{1}{y} \right)^2 
\right].
\label{eq:11} 
\end{equation}
where \( d\Omega = \sin \theta_{\text{CM}} \, d\theta_{\text{CM}} \, d\phi_{\text{CM}} \). Eq.~\eqref{eq:11} will be used in Section \ref{subsec:section4.3} to extract the scattering angle to determine the energy and momentum information of the scattered particles. The total cross section can be obtained by integrating Eq.~\eqref{eq:7} over \(y\) and $\phi_{\text{CM}}$

\begin{equation}
\begin{aligned}
\sigma &= \frac{4 r_e^2}{x^2} \int_{x / (x+1)}^{x} dy \int_{0}^{2\pi} d\phi_{\text{CM}} 
\left\{
\left( \frac{1}{x} - \frac{1}{y} \right)^2 
+ \left( \frac{1}{x} - \frac{1}{y} \right) 
+ \frac{1}{4} \left( \frac{x}{y} + \frac{y}{x} \right)
\right\} \\
&= \frac{2 \pi r_e^2}{x} 
\left\{
\left( 1 - \frac{4}{x} - \frac{8}{x^2} \right) \ln(1+x) 
+ \frac{1}{2} + \frac{8}{x} - \frac{1}{2(1+x)^2}
\right\}.
\label{eq:12} 
\end{aligned}
\end{equation}
The Klein-Nishina cross section is depicted in Fig.~\ref{Fig2}. The total cross section will be used in Section \ref{subsec:section4.2} to determine the number of event. When \(x \ll 1\), the scattering occurs in the Thomson regime, and the cross section asymptotically approaches the Thomson cross section  $\sigma_T \approx 6.652 \times 10^{-29} \, \text{m}^2$,  expressed as

\begin{equation}
\sigma = \frac{8 \pi r_e^2}{3} (1 - x) = \sigma_T (1 - x),
\label{eq:13} 
\end{equation}
The scattering cross section between electrons (\( E_e \approx 200~\mathrm{MeV} \)) and laser photons ($E_L \approx 1.18~\mathrm{eV}$), corresponding to $x \approx 3.6 \times 10^{-3}$, is $6.628 \times 10^{-29}$ $\mathrm{m^2}$.
When \(x \gg 1\), the scattering occurs in the Compton regime, and the cross section deviates significantly from \(\sigma_T\),  expressed as

\begin{equation}
\sigma = \frac{2 \pi r_e^2}{x} \left( \ln x + \frac{1}{2} \right).
\label{eq:14} 
\end{equation}
The scattering cross section between electrons (\( E_e \approx 200~\mathrm{MeV} \)) and gamma photons ($E_\gamma \approx 720~\mathrm{keV}$), corresponding to $x \approx 2205$, is $1.855 \times 10^{-31}$ $\mathrm{m^2}$.

\section{Results of inverse Compton scattering}\label{sec3}
This section presents the simulation results of two inverse Compton scattering processes. Particles with angles relative to the z-axis between $\pi/6$ and $5\pi/6$~rad will enter the detector and contribute to the background.
\subsection{The first scattering}\label{subsec:section3.3}

The beam parameters in Table~\ref{tab1} are used to simulate the process of the first inverse Compton scattering between electrons and laser photons. According to Eq.~\eqref{eq:8},  \( x \approx 3.6 \times 10^{-3} \ll 1 \), so the first inverse Compton scattering occurs in the Thomson regime. 

\begin{figure}[H]
\centering
\includegraphics[width=0.8\textwidth]{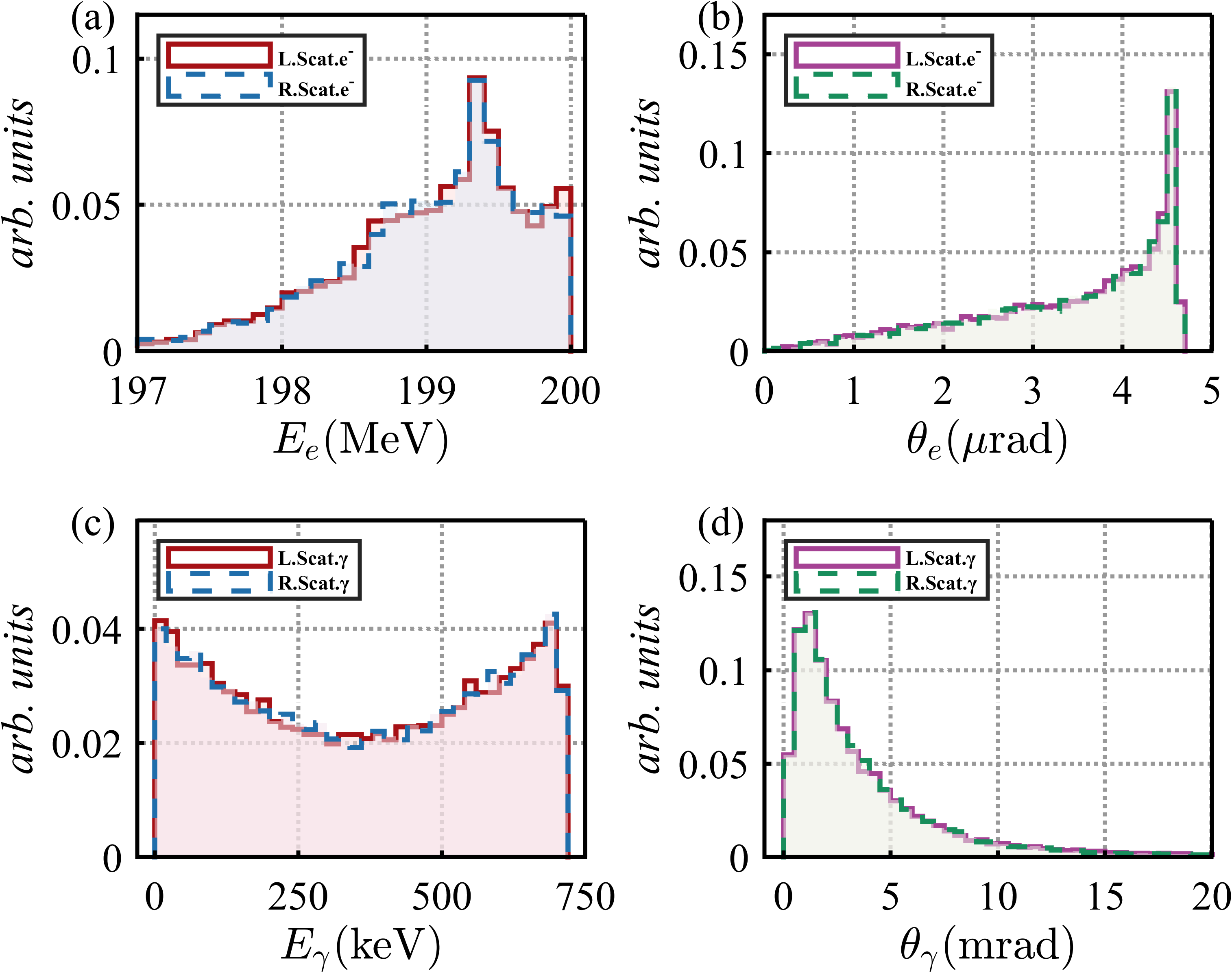} 

\caption{Energy and angular distribution of scattered particles on both sides after inverse Compton scattering (in the Thomson regime) of electrons (\( E_e\approx200~\mathrm{MeV} \)) and lasers (\( E_L\approx 1.18~\mathrm{eV} \)): (a) Electron energy spectrum. The electron energy is mainly concentrated around 199 MeV. Electrons lose hundreds of keV of energy in a single scattering, and some electrons will undergo multiple scatterings; (b) Electron scattering angle distribution. The electron scattering angle is in the order of $\mu$rad, so its propagation direction remains nearly unchanged; (c) Gamma photon energy spectrum. The energies of the gamma photons are on the order of 100~keV, with the Compton edge is about 720 keV. The overall energy spectrum shows a continuous wide spectrum distribution, with Compton photon yield on one side is approximately $1.48 \times 10^{12}$/s; (d) Gamma photon scattering angle distribution. The gamma photon scattering angle is on the order of mrad and is concentrated around 2.5 mrad. The (a) and (d) only show the main distribution areas of the data. $L$: Left; $R$: Right. This notation is used consistently throughout the paper.
}
\label{Fig3}
\end{figure}

\begin{figure}[H]
	\centering
	\includegraphics[width=0.8\textwidth]{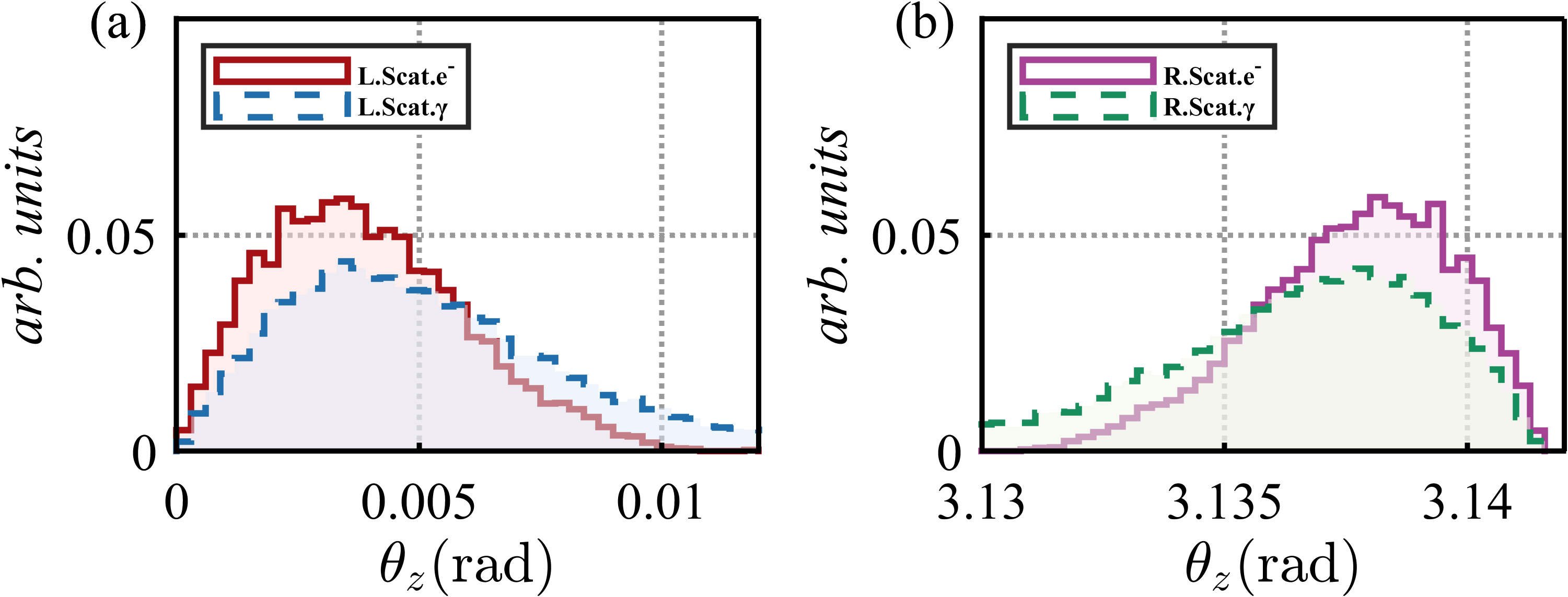}
	\caption{The angular distribution of scattered particles relative to the z-axis: Both electrons and gamma photons tend to be emitted in the forward direction.The peak angle relative to the electron’s direction of motion (±z axis) is 2.5 mrad. (a) and (b) show the main distribution areas of scattered electrons and scattered gamma photons on both sides.}
	\label{Fig4}
\end{figure}

The Compton photon yield on one side is approximately $1.48 \times 10^{12}$/s. As shown in Fig.~\ref{Fig3} (a) and Fig.~\ref{Fig3} (b), the scattering angle $\theta_e$ between the initial and final state electrons is very small, about the order of \text{urad}, losing a negligible amount of energy, with their propagation direction remaining almost unchanged and still maintains a straight trajectory during each scattering process. The energy of the scattered electrons is concentrated around 199 MeV, but can drop to as low as about 194 MeV because the fact that scattered electrons can undergo multiple collisions with laser photons, resulting in multiple scattering events. During the scattering process, the laser photons gain energy and are converted into gamma photons. Fig.~\ref{Fig3} (c) shows that the energy distribution of Compton photons presents a continuous wide energy spectrum, ranging from the energy of laser photons (1.18 eV) to the Compton edge 720 keV. From Fig.~\ref{Fig3} (d), it is evident that scattered photons tend to be emitted at small angles $\theta_\gamma$ relative to the initial electron direction. Specifically, approximately 1/2 of the gamma photons are emitted at an angle less than \( 1/\gamma = 2.5 \text{ mrad} \). Fig.~\ref{Fig4} shows that the angular distribution of scattered particles relative to the z-axis, denoted by $\theta_z$. Table~\ref{tab2} summarizes the energy range and number of background particles produced by two inverse Compton scatterings. Electrons from the first inverse Compton scattering do not enter the detector, while a small fraction of Compton photons with energies below 18 eV may enter the detector, contributing to background particles at a rate of $6.24 \times 10^7$/s. 

\begin{figure}[H]
	\centering
	\includegraphics[width=0.8\textwidth]{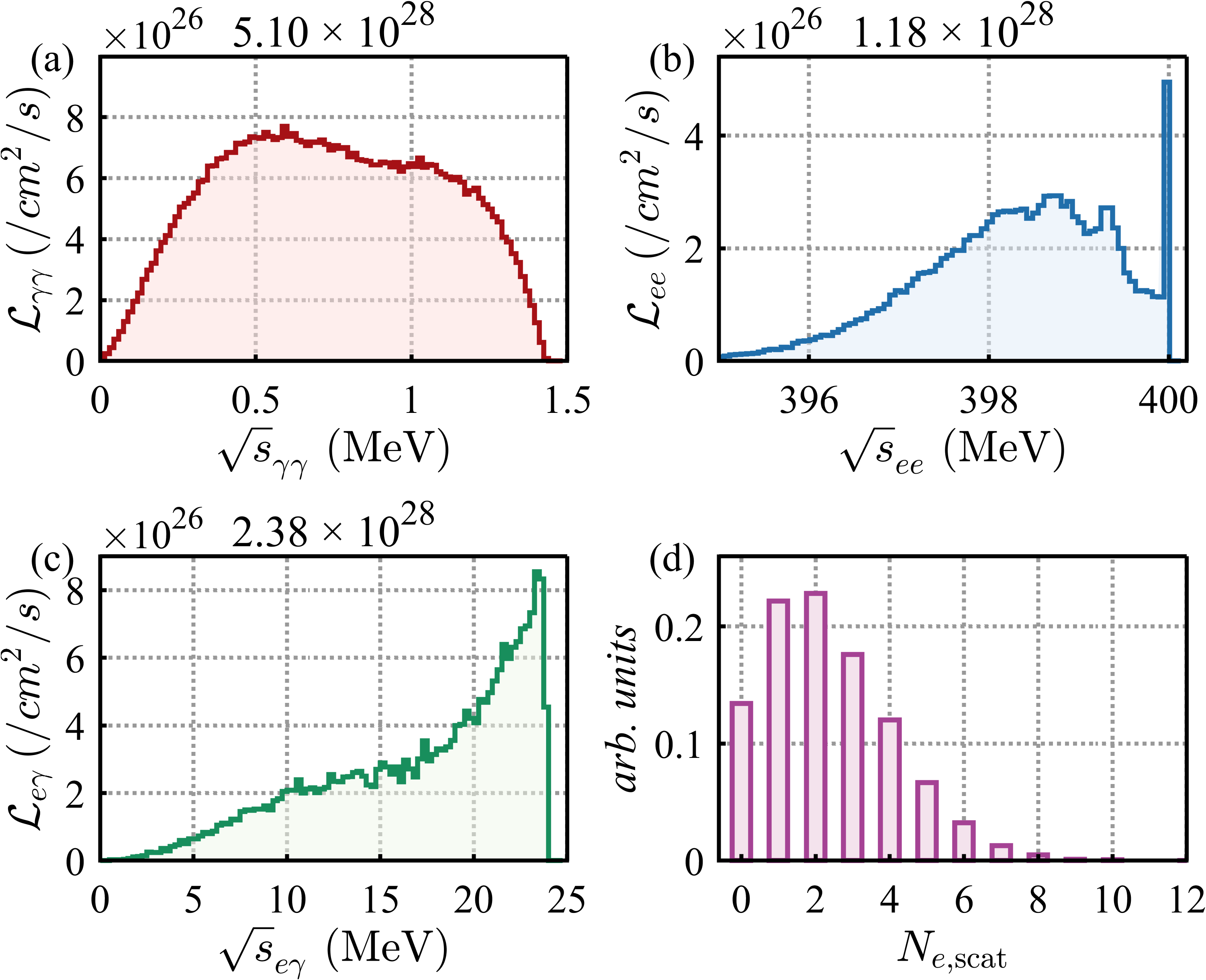}
	
	\caption{
		Luminosity distributions and electron scattering count after inverse Compton scattering.
		(a) Gamma-gamma luminosity $\mathcal{L}_{\gamma\gamma}$. The maximum CM energy of the Compton photons is approximately 1.5 MeV, and the total luminosity can reach \( 5.10 \times 10^{28} \, \mathrm{cm^{-2} \, s^{-1}} \); 
		(b) Electron-electron luminosity $\mathcal{L}_{ee}$. The CM energy of the electron pairs is concentrated above 396 MeV. The inverse Compton scattering process induces only small energy loss and negligible deflection in the electrons, thereby preserving the electron-electron luminosity at a high value of approximately \( 1.18 \times 10^{28} \, \mathrm{cm^{-2} \, s^{-1}} \);
		(c) Electron-gamma luminosity $\mathcal{L}_{e\gamma}$. The electron-gamma luminosity exhibits a peak near 25 MeV in the CM energy spectrum, with a total luminosity of \( 2.38 \times 10^{28} \, \mathrm{cm^{-2} \, s^{-1}} \);
		(d) Electron scattering count \( N_{\text{scat}} \). A substantial proportion of electrons undergo multiple scattering events, some even more than ten times.
	}
	\label{Fig5} 
\end{figure}

Fig.~\ref{Fig5} presents the differential luminosity and electron scattering count after inverse Compton scattering. Fig.~\ref{Fig5} (a) displays the gamma-gamma luminosity, achieving a total value of \( 5.10 \times 10^{28} \, \mathrm{cm^{-2} \, s^{-1}} \), which signifies highly efficient photon generation. Fig.~\ref{Fig5} (b) shows the electron-electron luminosity maintained at a high level of \( 1.18 \times 10^{28} \, \mathrm{cm^{-2} \, s^{-1}} \), with CM energies above 396~MeV. This demonstrates that the electron beam quality is well maintained during the inverse Compton scattering. Fig.~\ref{Fig5} (c) exhibits the electron-gamma luminosity, showing a characteristic peak near 25~MeV and a total luminosity of \( 2.38 \times 10^{28} \, \mathrm{cm^{-2} \, s^{-1}} \). Fig.~\ref{Fig5} (d) shows a substantial number of electrons experience multiple scattering, with some undergoing more than ten events. This high multiplicity provides direct evidence of the dense interaction environment under high luminosity.

\subsection{The second scattering}\label{subsec:section3.2}

The second inverse Compton scattering occurs between electrons and gamma photons. As can be seen from Fig.~\ref{Fig4} (c), the energies of the generated gamma photons are on the order of $100$ \text{keV}. As shown in Fig.~\ref{Fig2}, the cross section of the second inverse Compton scattering can drop to $1.855 \times 10^{-31}$ $\mathrm{m^2}$, which is approximately three orders of magnitude smaller than the cross section of the first inverse Compton scattering. According to Eq.~\eqref{eq:8}, \( x \gg 1 \), so the second inverse Compton scattering occurs in the Compton regime \cite{27}. 

\begin{figure}[H]
\centering
\includegraphics[width=0.8\textwidth]{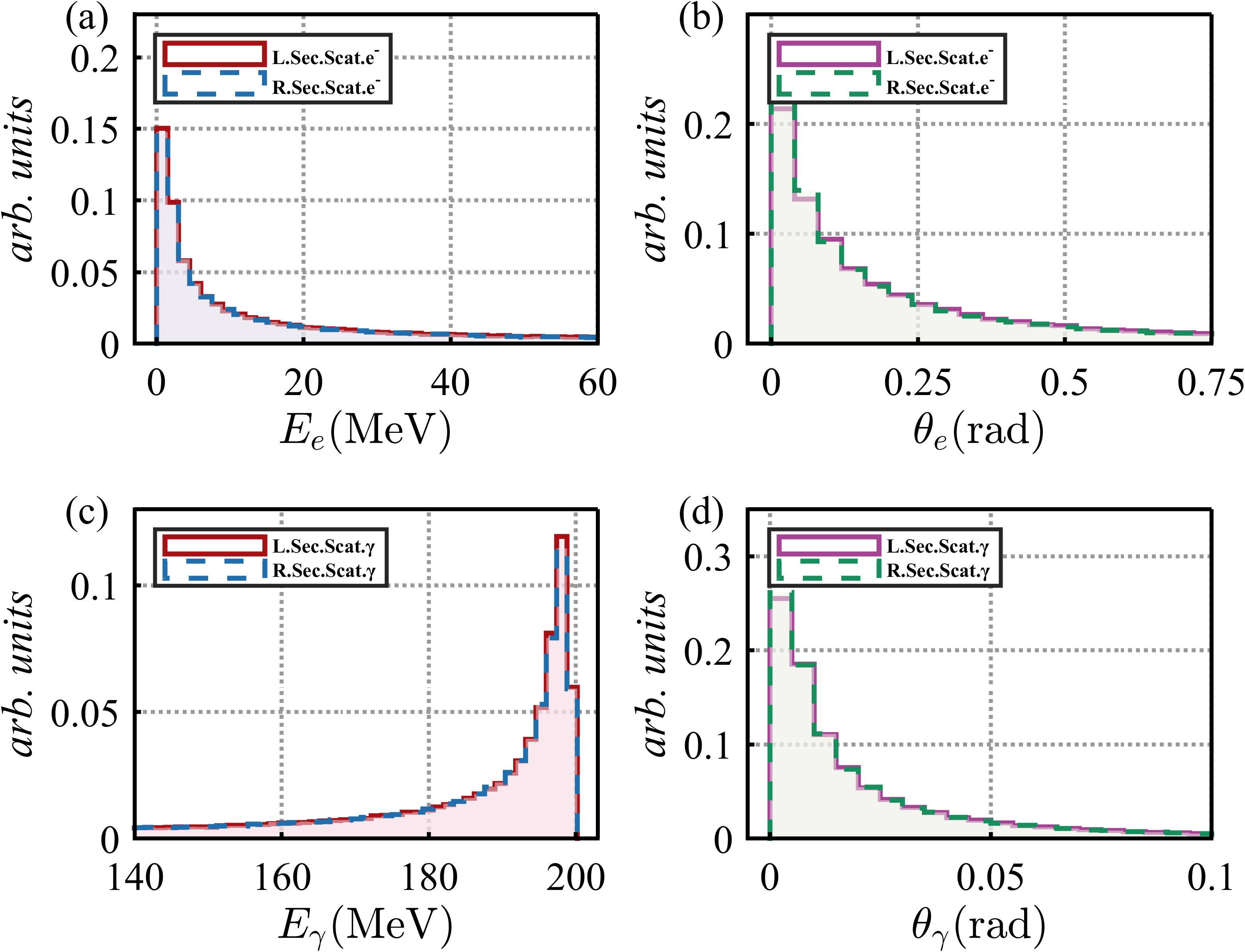}
\caption{Energy and angular distribution of scattered particles on both sides after second Compton scattering (in the Compton regime) of electrons (\( E_e\approx200~\mathrm{MeV} \)) and gamma photons ( $E_\gamma$ is on the order of $100~\mathrm{keV}$), the total number of second inverse Compton events on both sides is approximately 245.77/s: (a) Energy spectrum distribution of electrons. The scattered electron energy spectrum covers a wide range from 0.511 MeV to 200 MeV, with a concentration in the low-energy region. (b) Scattering angle distribution of scattered electrons. The electron experiences a relatively large deflection in this process, reaching the rad scale, while the majority are concentrated within 0.08 rad. (c) Energy spectrum distribution of secondary gamma photons. The energy of the scattered photons ranges from tens of keV to 200 MeV, with a concentration in the high-energy region; (d) Scattering angle distribution of secondary gamma photons. The scattering angle reaches the rad level, mainly concentrated within 0.01 rad, and is overall smaller than that of electrons. The graph only show the main distribution areas of the data.}
\label{Fig6}
\end{figure}

\begin{figure}[H]
\centering
\includegraphics[width=0.8\textwidth]{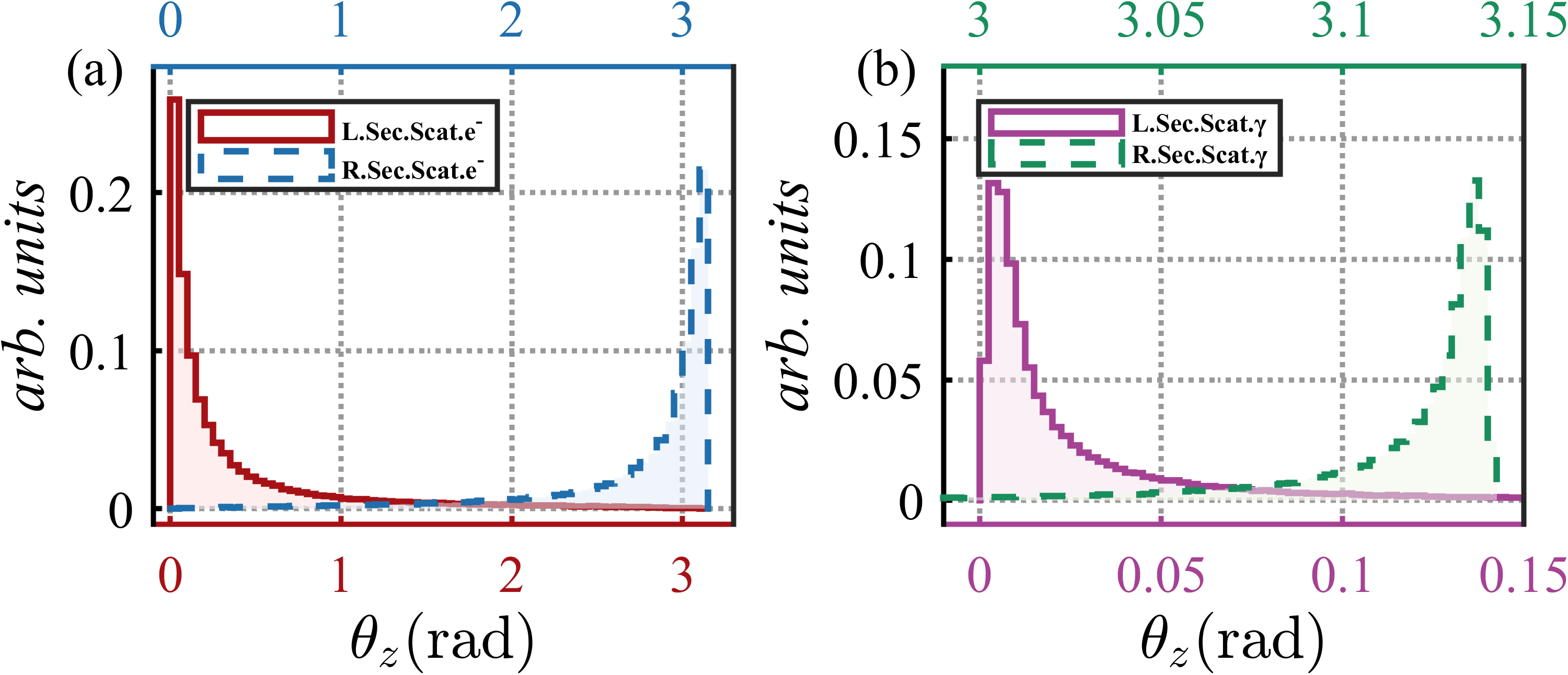}
\caption{The angular distribution of scattered particles relative to the z-axis: (a) The angle between the scattered electrons and the z-axis. The angle distribution ranges from 0 to $\pi$ rad, clearly indicating that reverse electrons appear in this process. (b) The angle between the scattered photons and the z-axis. Compared with electrons, the angular distribution of gamma photons is more concentrated, indicating a stronger forward orientation and smaller angular divergence.}
\label{Fig7}
\end{figure}

The total number of second inverse Compton events on both sides is approximately 245.77/s. Fig.~\ref{Fig6}(a) shows the energy spectrum of scattered electrons, and Fig.~\ref{Fig6}(b) shows their scattering angles. The energy distribution of the scattered electrons ranges from 0.511 keV to 200 MeV, and only the main part is shown. The energy spectrum of the scattered electrons shows a concentration in the low-energy region, indicating that electrons tend to transfer all its kinetic energy to the photons with a larger scattering angle during the scattering process. The electron recoil cannot be neglected, which is in agreement with Ref. \cite{28}. Fig.~\ref{Fig6}(c) shows the energy spectrum of scattered gamma photons, and Fig.~\ref{Fig6}(d) shows their scattering angles. The energy of the scattered photons ranges from tens of keV to approximately 200 MeV, with a concentration in the high-energy region. The maximum energy of the scattered photons is comparable to the initial electron energy but is constrained by the kinematic limit \( E_e \). In this process, the scattering angle of electrons is greater than the scattering angle of gamma photons overall. Fig.~\ref{Fig7} (a) and Fig.~\ref{Fig7} (b) show the angle between the scattered particles and the z-axis. It is noteworthy that a portion of the scattered electrons moving in the reverse direction. As shown in Table~\ref{tab2}, the background electron count rate is approximately 51.99/s, while the background gamma photon count rate is 0.99/s. The energies of both are below 11 MeV.

\subsection{Background of other process}\label{subsec:section3.3}

\subsubsection{Background from Møller Scattering}

In addition, although this study primarily focuses on background particles generated by the inverse Compton scattering process, we also performed a preliminary estimation of background electrons arising from Møller scattering, motivated by concerns regarding its potential to generate high-energy electrons that may affect detector performance. The results show that under the current parameter conditions, electrons with an energy of about 200 MeV will enter the detector at a rate of 0.56/s. The potential impact of Møller scattering background on detector design should be carefully taken into account.

\subsubsection{Background from the Breit-Wheeler Process}
So far, the Breit-Wheeler process of two real photons has not been directly observed. The first experimental verification of this process is expected to be achieved with the MeV Gamma-Gamma Collider. Based on the parameters listed in Table~\ref{tab1}, the maximum energy of gamma photons produced in the first inverse Compton scattering can reach up to 720 keV (the energy in the CM of the two Compton photons being\(\sqrt{s_{\gamma\gamma}} \lesssim 1.44~\text{MeV}\)). This exceeds the Breit-Wheeler threshold ($\sqrt{s} = 2m_e c^2 = 1.022\,\text{MeV}$), enabling electron-positron pair production. The count rates of electrons and positrons entering the detector are 1312.2/s and 1314.3/s, respectively, with their energy distributions ranging from 511 to 720 keV.

\begin{figure}[H]
\centering
\includegraphics[width=0.8\textwidth]{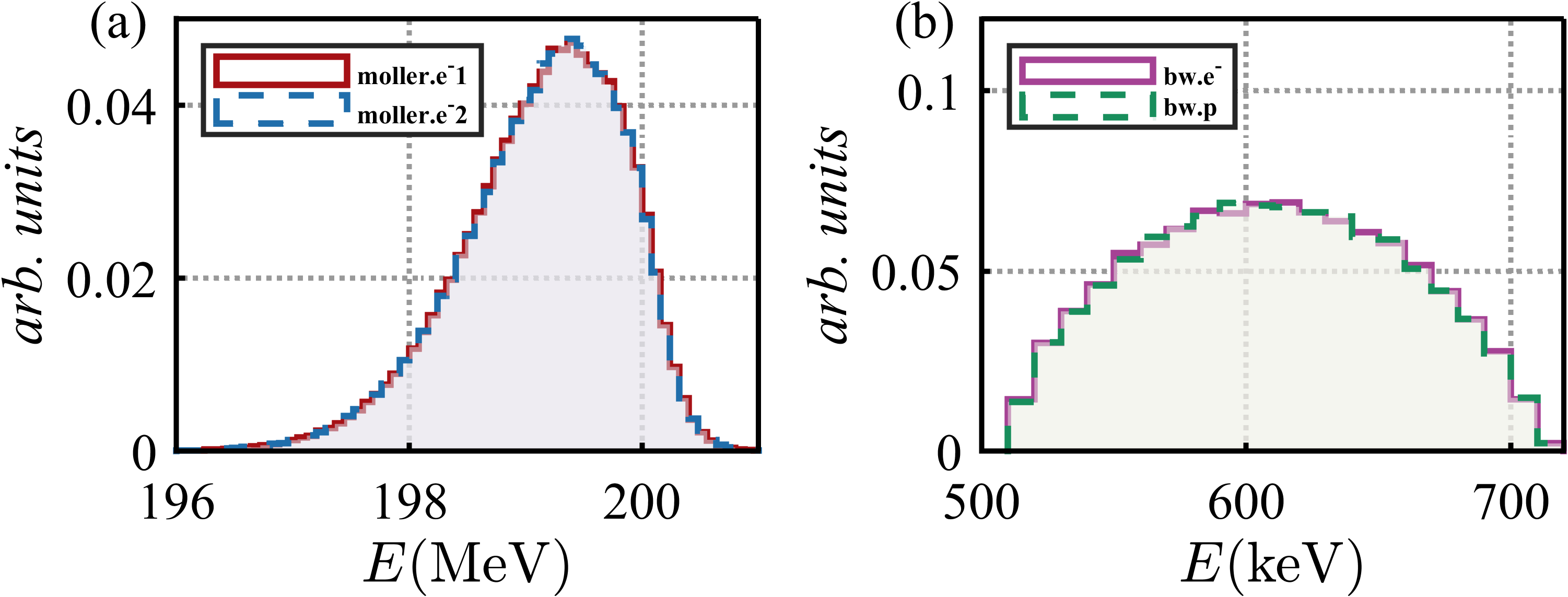} 

\caption{Energy spectrum of scattered particles in Møller and Breit-Wheeler scattering: (a) Energy spectrum of scattered particles from Møller scattering. The electron energy spectrum is approximately a Gaussian distribution around 199 MeV; (b) Energy spectrum of scattered particles from Breit-Wheeler scattering. The energy distribution range of positrons (bw.p) and electrons ($\mathrm{bw}.\mathrm{e}^{-}$)
 is from 511 to 720 keV.}
\label{Fig8}
\end{figure}
\renewcommand{\arraystretch}{1.25}
\begin{table}[htbp!]
	\centering
	\caption{\centering Background particles: number and energy distribution} \label{tab2}%
	
	\noindent
	\begin{tabularx}{\textwidth}{
			>{\centering\arraybackslash\hsize=1\hsize}X
			>{\centering\arraybackslash\hsize=0.5\hsize}X
			>{\centering\arraybackslash\hsize=0.5\hsize}X
			>{\centering\arraybackslash\hsize=0.5\hsize}X
			>{\centering\arraybackslash\hsize=0.5\hsize}X
		}
		\toprule
		Scattering Process & $N_{e}$/s & $E_{e}$/MeV & $N_{\gamma}$/s & $E_{\gamma}$/MeV\\
		\midrule
		First Inverse Compton & / & / & $6.24 \times 10^7$ & \( < 1.8 \times 10^{-5} \) \\
		Second Inverse Compton & 51.99 & \( < 11 \) & 0.99 & \( < 11 \)  \\
		Møller & 0.56 & \( \sim 200 \) & / & / \\
		Breit-Wheeler & 2626.5 & \( 0.511 - 0.72 \) & / & / \\
		\bottomrule
		\multicolumn{5}{@{}p{\textwidth}@{}}{Background particles are defined as those with an angle of \( \pi/6 \) to \( 5\pi/6 \) radians to the z-axis.}
	\end{tabularx}
\end{table}
\section{Methods}\label{sec4}

\begin{figure}[h]
\centering
\includegraphics[width=1\textwidth]{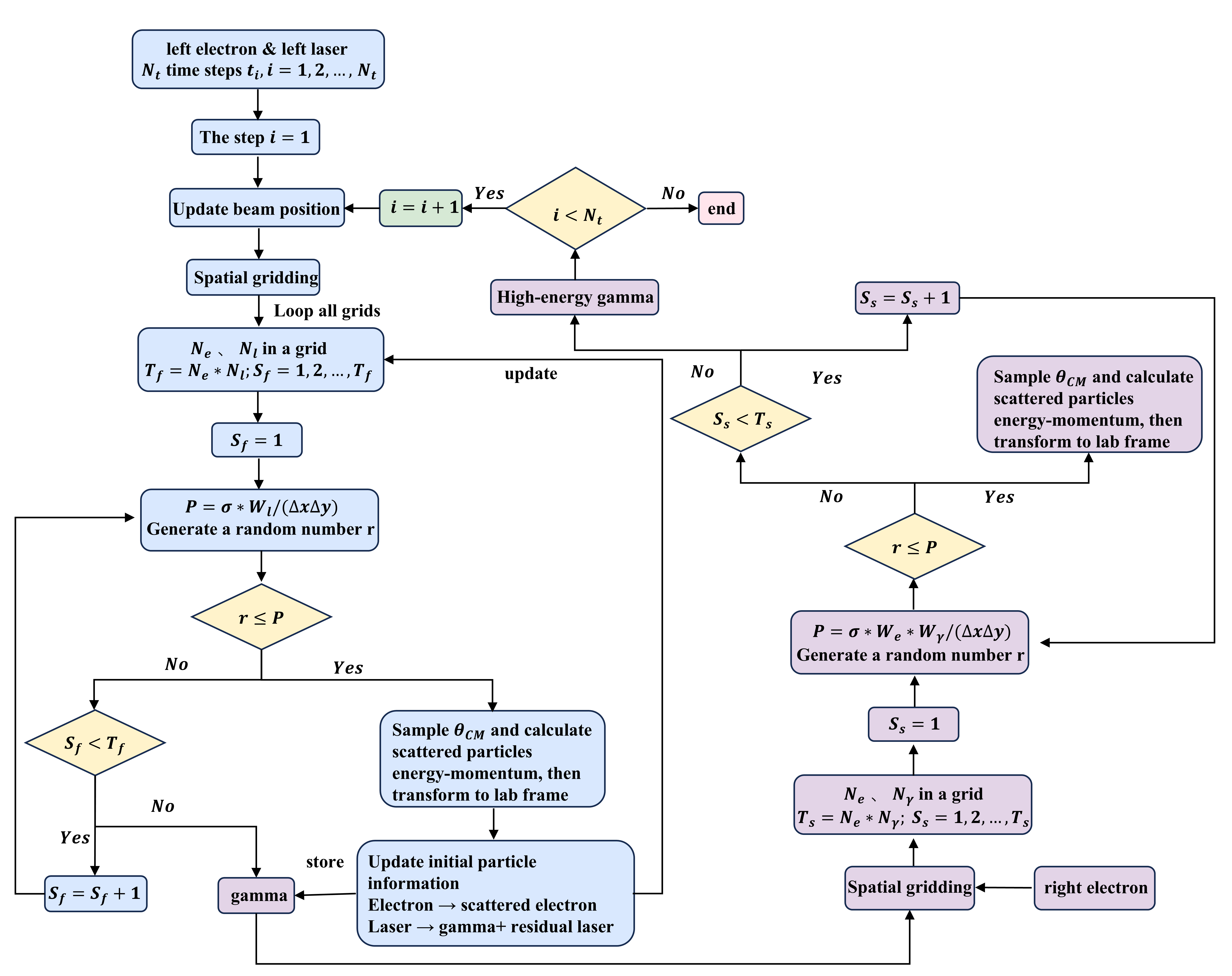}
\caption{Schematic of the GBET simulation workflow. This flowchart illustrates the algorithm for simulating two sequential inverse Compton scattering events in GBET. The process begins with the initialization of electron and laser beam parameters. The entire interaction is discretized into \( N_t \) time steps. Within each step, the beam positions are first updated. All electron-photon pairs within the same cell are iteratively evaluated for scattering. Each scattering event leads to the replacement of the initial electron with the scattered one and the splitting of the laser photon into a new macro-gamma photon and a residual one with a reduced weight. All three particles are returned to the particle pool for subsequent interactions. After all electron-photon pairs within the time step are processed, the macro-gamma photon beam is subsequently checked for scattering with the counter-propagating electron beam. Owing to the sparse scattering probability, the incident macro-particles are preserved intact and returned to the particle pool, with the real particle products being stored as output and excluded from the subsequent cycle. The above steps are repeated until all time steps are completed. The right-side reactions are identical to those on the left and are therefore omitted for clarity.}\label{Fig9}
\end{figure}

Based on MC techniques, this paper takes advantage of the intrinsic randomness of QED processes to develop a code, GBET, that can handle macro-particles of different weights and maintain momentum and energy conservation in each scattering. The core advantage of GBET is its ability to perform an integrated simulation of two consecutive inverse Compton scattering processes. This contrasts with traditional chain-simulation methods based on luminosity spectra, which lose critical phase-space information; by preserving complete particle-level information, GBET achieves higher physical accuracy. In addition, GBET improves computational efficiency through parallel processing. The effectiveness of the code is verified by benchmarking against the simulation results of CAIN and the theoretical integration results.  The structure of the GBET is shown in Fig.~\ref{Fig9}.

\subsection{Spatial grid construction}\label{subsec:section4.1}

At the particle level, tracking each particle in the simulation is impractical due to the large number of particles in the beam. Therefore, the beam is divided into weighted macro-particles. Temporally, the beam-beam interaction dynamics is discretized into $N$ time steps of size $\Delta t$. During each time step, the particle's position is evolved according to its equation of motion. Spatially, assuming that collisions require spatial proximity, the electron region is divided into a three-dimensional space grid, and only particle pairs in the same grid cell are considered for collisions~\cite{30}. A spatial grid with a uniform transverse resolution of $\Delta x = \Delta y = 0.27\,\mu\text{m}$ is used in the simulation. 
The longitudinal grid spacing along the collision axis is determined by
\begin{equation}
\Delta z = \left| \vec{v}_e - \vec{v}_{\text{ph}} \right| \Delta t = c \left| 1 - \beta \cos{\theta_i} \right| \Delta t,
\label{eq:15}
\end{equation}
where \( \vec{v}_e \) and \( \vec{v}_{\text{ph}} \) are the velocities of the electron and photon, respectively, c is the speed of light, \( \beta = v_e / c \), and \( \theta_i \) is the collision angle. This grid spacing corresponds to the overlap length of the two beams over the time interval $\Delta t$.
\subsection{ Scattering probability}\label{subsec:section4.2}

The number of potential scattering pairs within a grid cell containing $N_e$ macro-electrons and $N_{\text{photon}}$ macro-photons is $S = N_e \times N_{\text{photon}}$, with each pair having a probability of interaction. For each electron-photon pairs, the scattering events is calculated using Eq.~\eqref{eq:16}.
\begin{equation}
N_{phys} = \sigma_{\text{tot}} \mathcal{L}, \quad
\mathcal{L} = \frac{w_e w_{ph}}{\Delta x \, \Delta y}
\label{eq:16}
\end{equation}
 where the total cross section $\sigma_{\text{tot}}$ depends on the momenta of the incident electron and photon, \( w_e \) and \( w_{ph} \) denote the weights of the incident electron and photon, respectively. The number of physical scattering events, \( N_{\text{phys}} \), given by Eq.~\eqref{eq:16} is related to the number of simulation events by \( N_{\text{sim}} = N_{\text{phys}} / w \), where \( w \) is the weight of the generated macro-particle~\cite{31}. When $N_{\text{sim}} \ll 1$, the event probability satisfies $P \approx N_{\text{sim}}$. Whether scattering occurs is determined by the rejection sampling method~\cite{31,32,33}. Generate a random number \(r\) uniformly in the range of 0 to 1. If \(r \leq P\), the scattering occurs and the process proceeds to the third step; otherwise, reselect the macro-particle pairs and repeat the second step~\cite{34}. 

\begin{figure}[H]
\centering
\includegraphics[width=0.8\textwidth]{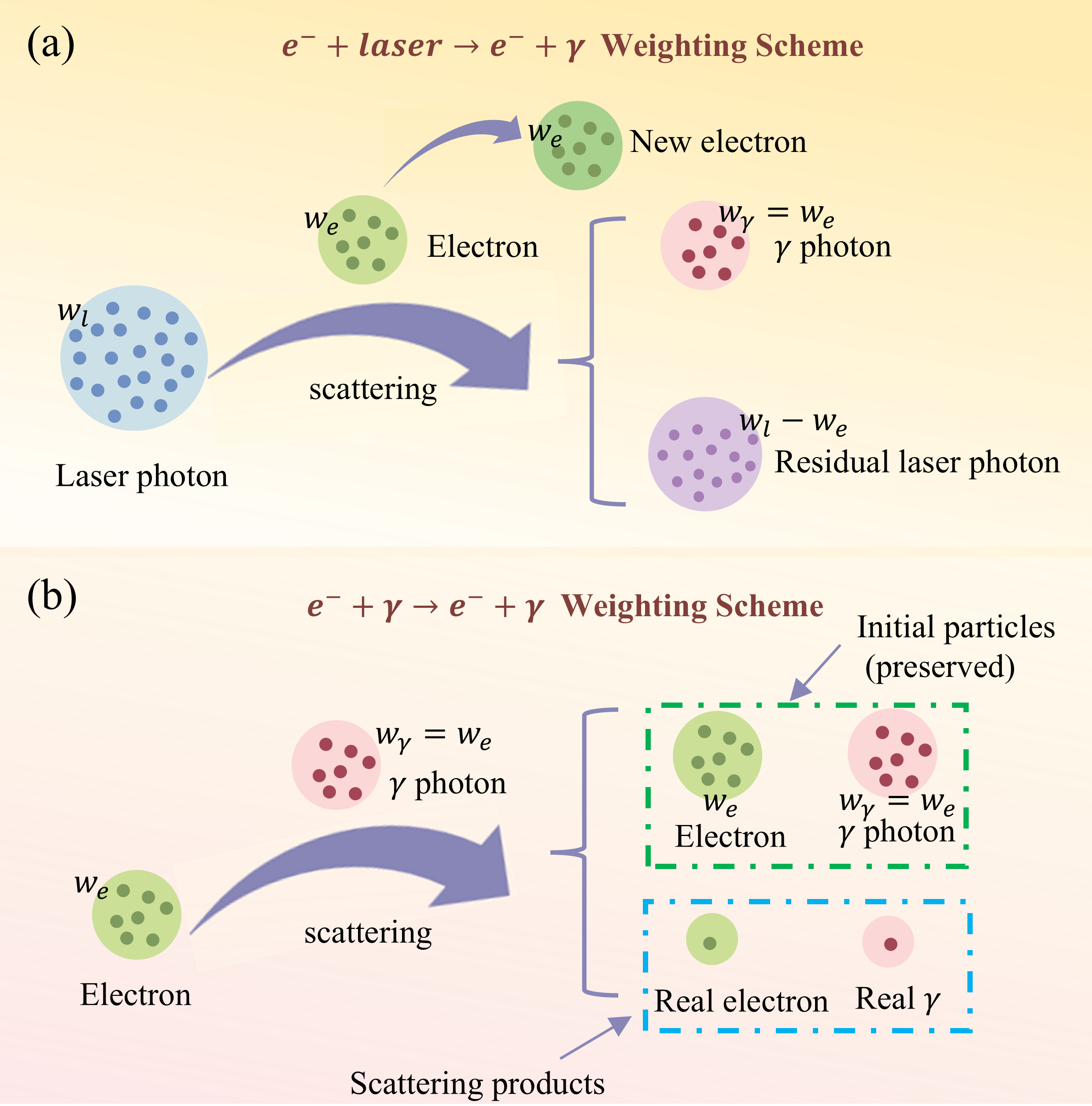}
\caption{Schematic diagram of weight processing for two inverse Compton scattering processes in different regimes in the MeV Gamma-Gamma Collider: (a) A macro-electron with weight $w_e$ interacts with a macro-laser photon of weight $w_l$. The laser photon splits, generating a new macro-gamma photon with a weight $w_\gamma = w_e$, while the residual macro-laser photon retains all its initial properties except for the weight, which is reduced to $w_l - w_e$. The initial electron is replaced by the scattered one. All three particles re-enter the particle pool to participate in subsequent reactions.
	 (b) A macro-electron collide with a macro-gamma photon. Due to the extremely low scattering probability, the interactions between real particles are exceedingly sparse. Consequently, the products are generated as real particles (real electrons and real $\gamma$ photons), while the impact on the initial macro-particle is negligible. The initial macro-electron and macro-gamma photon are returned to the particle pool to continue reactions.}\label{Fig10}
\end{figure}

In the simulation, the two inverse Compton scattering processes are implemented using different weighting schemes. As shown in Fig.~\ref{Fig10} (a), in simulations of inverse Compton scattering between laser photons and electrons, the interacting macro-electron and macro-photon have unequal weights, \(w_e\) and \(w_l\), respectively. When a scattering event is triggered, all $w_e$ real electrons in the macro-electron are made to scatter with $w_e$ real photons in the macro-photon. After scattering: (1) the incident macro-electron is replaced by a new one with energy and momentum updated according to Compton kinematics; (2) the incident macro-photon splits into two components: a scattered macro-gamma photon with weight \(w_\gamma = w_e\) (carrying the updated momentum) and a residual macro-photon with weight \(w_l' = w_l - w_e\) (preserving its initial kinematic state, with only the weight being altered). These particles are returned to the particle pool, where they continue to pair up with other particles and possibly undergo multiple scatterings.

As shown in Fig.~\ref{Fig10} (b), in simulations of inverse Compton scattering between gamma photons and electrons, both have identical weights (\(w_\gamma = w_e\)). Owing to the extremely low scattering probability, interactions among the real particles are exceedingly sparse, resulting in negligible impact on the initial macro-particle. Therefore, the incident macro-particles are preserved, and the scattering products are modeled as individual real particles, with their energy and momentum determined by sampling the scattering angle and applying energy-momentum conservation. This processing method accurately reflects the sparsity of the physical process. In addition, sufficient statistics can be obtained by employing an artificially enhanced cross section.

\subsection{Scattering product kinematics}\label{subsec:section4.3}

For each triggered scattering event in the simulation, the scattered particle properties are determined in the CM frame. The total CM energy of the binary reaction process ($12 \rightarrow 34$):
\begin{equation}
\sqrt{s} = \sqrt{m_1^2 + m_2^2 + 2(E_1 E_2 - \vec{p}_1 \cdot \vec{p}_2)},
\label{eq:17}
\end{equation}
where \( E_1 \), \( E_2 \) and \( \vec{p}_1 \), \( \vec{p}_2 \) are the energies and momenta of the incoming particles in the laboratory frame.
In the CM system, the energies of the final-state particles are
\begin{equation}
E_{3,4}^{\mathrm{CM}} = \frac{s + m_{3,4}^2 - m_{4,3}^2}{2\sqrt{s}}
\label{eq:18},
\end{equation}
and the momenta satisfy \( p^{\mathrm{CM}}_3 = - p^{\mathrm{CM}}_4 \).
The components of the scattered particles’ momenta are

\begin{equation}
\begin{aligned}
p_{3x}^{\mathrm{CM}} &= p_3^{\mathrm{CM}} \sin \theta_{\mathrm{CM}} \cos \phi_{\mathrm{CM}}, \quad
p_{3y}^{\mathrm{CM}} = p_3^{\mathrm{CM}} \sin \theta_{\mathrm{CM}} \sin \phi_{\mathrm{CM}}, \quad
p_{3z}^{\mathrm{CM}} = p_3^{\mathrm{CM}} \cos \theta_{\mathrm{CM}}, \\
p_{4x}^{\mathrm{CM}} &= p_4^{\mathrm{CM}} \sin \theta_{\mathrm{CM}} \cos \phi_{\mathrm{CM}}, \quad
p_{4y}^{\mathrm{CM}} = p_4^{\mathrm{CM}} \sin \theta_{\mathrm{CM}} \sin \phi_{\mathrm{CM}}, \quad
p_{4z}^{\mathrm{CM}} = p_4^{\mathrm{CM}} \cos \theta_{\mathrm{CM}}.
\end{aligned}
\label{eq:19}
\end{equation}
where $\theta_{\mathrm{CM}}$ and $\phi_{\mathrm{CM}}$ are the scattering angle and azimuth angle, respectively. Then, the inverse Lorentz transformation is carried out back to the laboratory system. Given the four-momentum in the CM frame as $ p_{\text{CM}}^\mu = (E_{\text{CM}}, p_{x,\text{CM}}, p_{y,\text{CM}}, p_{z,\text{CM}}) $, we can transform it to the laboratory frame using:
\begin{equation}
p_{\text{Lab}}^\mu = L \cdot p_{\text{CM}}^\mu
\label{eq:20}
\end{equation}
$L$  is given by 
\begin{equation}
L = \begin{pmatrix}
\gamma_{\text{CM}} & \beta_{x,\text{CM}} \gamma_{\text{CM}} & \beta_{y,\text{CM}} \gamma_{\text{CM}} & \beta_{z,\text{CM}} \gamma_{\text{CM}} \\
\beta_{x,\text{CM}} \gamma_{\text{CM}} & 1 + (\gamma_{\text{CM}} - 1) \frac{\beta_{x,\text{CM}}^2}{\beta_{\text{CM}}^2} & (\gamma_{\text{CM}} - 1) \frac{\beta_{x,\text{CM}} \beta_{y,\text{CM}}}{\beta_{\text{CM}}^2} & (\gamma_{\text{CM}} - 1) \frac{\beta_{x,\text{CM}} \beta_{z,\text{CM}}}{\beta_{\text{CM}}^2} \\
\beta_{y,\text{CM}} \gamma_{\text{CM}} & (\gamma_{\text{CM}} - 1) \frac{\beta_{x,\text{CM}} \beta_{y,\text{CM}}}{\beta_{\text{CM}}^2} & 1 + (\gamma_{\text{CM}} - 1) \frac{\beta_{y,\text{CM}}^2}{\beta_{\text{CM}}^2} & (\gamma_{\text{CM}} - 1) \frac{\beta_{y,\text{CM}} \beta_{z,\text{CM}}}{\beta_{\text{CM}}^2} \\
\beta_{z,\text{CM}} \gamma_{\text{CM}} & (\gamma_{\text{CM}} - 1) \frac{\beta_{x,\text{CM}} \beta_{z,\text{CM}}}{\beta_{\text{CM}}^2} & (\gamma_{\text{CM}} - 1) \frac{\beta_{y,\text{CM}} \beta_{z,\text{CM}}}{\beta_{\text{CM}}^2} & 1 + (\gamma_{\text{CM}} - 1) \frac{\beta_{z,\text{CM}}^2}{\beta_{\text{CM}}^2}
\end{pmatrix}
\label{eq:21}
\end{equation}
where \( \gamma_{\mathrm{CM}} = (E_1 + E_2)/\sqrt{s} \) and \( \beta_{\mathrm{CM}} = (\beta_{x,\mathrm{CM}}, \beta_{y,\mathrm{CM}}, \beta_{z,\mathrm{CM}})=(p_1 + p_2)/(E_1 + E_2) \). Therefore, to determine the energy and momentum of the scattered particles, it is necessary to first determine the scattering angle and azimuth angle. In the unpolarized case, the differential cross section exhibits azimuthal symmetry, allowing the azimuthal angle $\phi_{\text{CM}}$ to be uniformly sampled in the interval \([0, 2\pi)\). By integrating Eq.~\eqref{eq:11} over the azimuthal angle $\phi_{\text{CM}}$, we obtain:

\begin{equation}
\frac{d\sigma}{d\theta_{\mathrm{CM}}} = \frac{4\pi r_e^2 (mc^2)^2 \sin\theta_{\mathrm{CM}}}{s} \left[ 
\frac{1}{4} \left( \frac{x}{y} + \frac{y}{x} \right) + 
\left( \frac{1}{x} - \frac{1}{y} \right) + 
\left( \frac{1}{x} - \frac{1}{y} \right)^2 
\right].
\label{eq:22} 
\end{equation}
The scattering angles $\theta_{\text{CM}}$ are sampled in the CM frame according to the differential cross section, using the inverse transform sampling method ~\cite{35,36}. Subsequently, the energy and momentum of the scattered particles are updated according to Eqs.~\eqref{eq:17}--~\eqref{eq:21}. The algorithm then proceeds to the next electron-photon pair and repeats the above process. After all particle pairs have been processed, the beam is advanced by a single time step $\Delta t$, and the entire procedure is repeated.

\subsection{Benchmarking for Inverse Compton Scattering}\label{subsec:section4.4}
Based on the above algorithm, the GBET code has been developed to fully simulate two successive inverse Compton scattering processes in Gamma-Gamma Collider. We briefly present the benchmarking results of the code below, validated against theoretical predictions and CAIN simulations. The benchmark scenario involves the head-on collision of two electron beams and two laser pulses, simulated using the parameters given in Table ~\ref{tab1}.

\subsubsection{Theoretical Predictions}\label{subsec:section4.4.1}
The luminosity for the pulse-pulse collision case with a crossing angle $\Phi$ is ~\cite{37,38} :
\begin{equation}
\begin{split}
\mathcal{L} &= \int n_1 n_2 \cdot 2c \cdot \cos(\Phi) \, dV \\
            &= \int N_1 f_1(x_1, y_1, z_1, t) \, N_2 f_2(x_2, y_2, z_2, t) \cdot 2cBF \cdot \cos^2(\Phi) \, dx \, dy \, dz \, dt
\end{split}
\label{eq:23}
\end{equation}
Here, \( c \) denotes the speed of light, \( dV \) is the overlap volume of the two colliding bunches, and \( n_1 \) and \( n_2 \) are the particle densities of beam~1 and beam~2, respectively. For pulsed beams, the particle density is given by \( n_i = N_i f_i(x_i, y_i, z_i, t) \), where \( N_i \) denotes the total number of particles in beam \( i \), and the distribution function \( f_i \) is normalized such that
\( \int f_i(x_i, y_i, z_i, t) \, dx_i \, dy_i \, dz_i = 1 \). \( BF \) represents the number of collisions per unit time, where \( B \) is the number of bunches and \( F \) is the revolution frequency. The distribution functions of the pulsed electron and laser beams, both modeled as three-dimensional Gaussian distributions, are expressed as follows:
\begin{equation}
\begin{split}
f_e(e_x, e_y, e_z, t) &= \frac{1}{(2\pi)^{\frac{3}{2}}} \frac{1}{\sigma_{e_x} \sigma_{e_y} \sigma_{e_z}} \exp \left[ -\frac{1}{2} \left( \frac{x_e^2}{\sigma_{e_x}^2} + \frac{y_e^2}{\sigma_{e_y}^2} + \frac{(z_e - ct)^2}{\sigma_{e_z}^2} \right) \right] \\
f_l(l_x, l_y, l_z, t) &= \frac{1}{(2\pi)^{\frac{3}{2}}} \frac{1}{\sigma_{l_x} \sigma_{l_y} \sigma_{l_z}} \exp \left[ -\frac{1}{2} \left( \frac{x_l^2}{\sigma_{l_x}^2} + \frac{y_l^2}{\sigma_{l_y}^2} + \frac{(z_l + ct)^2}{\sigma_{l_z}^2} \right) \right]
\end{split}
\label{eq:24}
\end{equation}
By applying the parameters from Section~\ref{subsec:section4.4} to Eq.~\eqref{eq:23}, the luminosity is calculated to be \(2.314 \times 10^{36}~\mathrm{cm}^{-2}\mathrm{s}^{-1}\). Given $ x \approx 3.6 \times 10^{-3} \ll 1 $ from Eq.~\eqref{eq:8}, the total cross section from Eq.~\eqref{eq:14} is \(6.628 \times 10^{-29}\ \mathrm{m}^2\). Consequently, the total number of scattering events is estimated as \(1.534 \times 10^{12}\)/s.

\subsubsection{Comparison with CAIN}\label{subsec:section4.4.2}

\begin{figure}[H]
\centering
\includegraphics[width=0.8\textwidth]{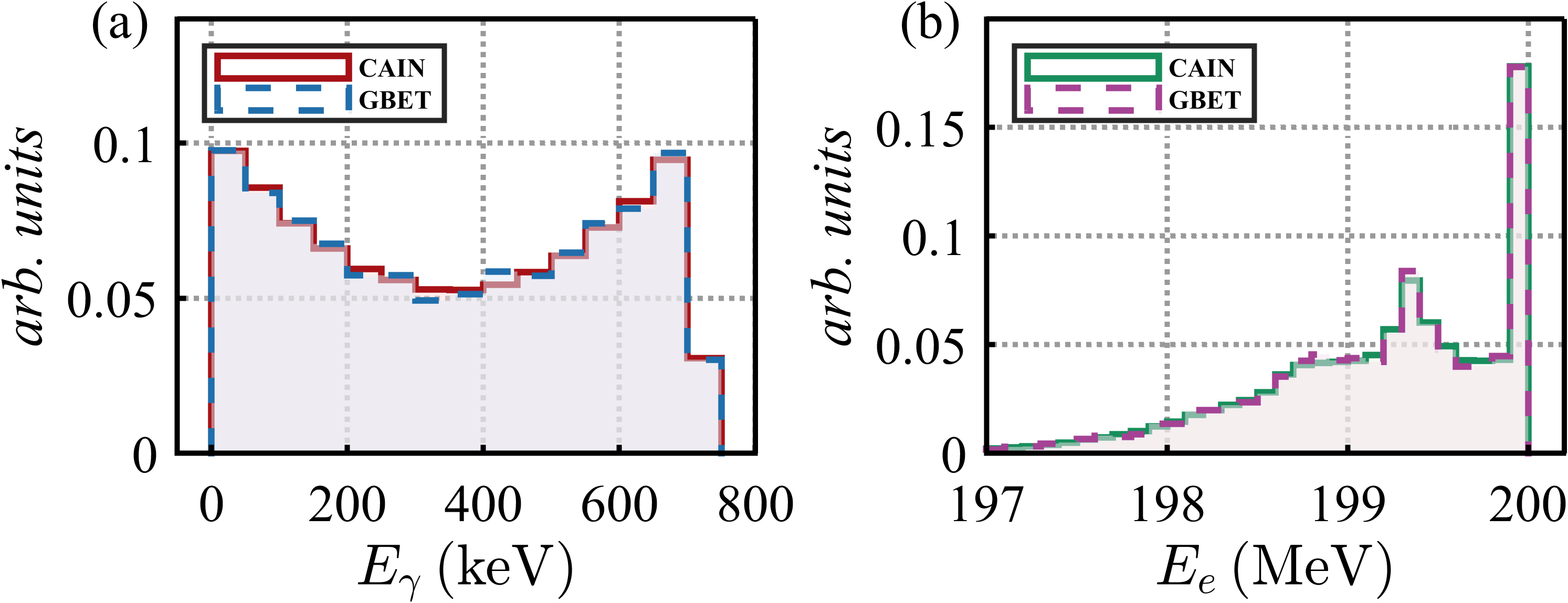}
\caption{Comparison of the energy spectra between GBET and CAIN. The Compton photon energy spectrum is shown in panel (a), and the spectra of all electrons after Scattering in panel (b).}
\label{Fig11}
\end{figure}

 Table \ref{tab3} compares the key output quantities from the GBET and CAIN simulations, including the Compton photon yield \(N_\gamma\), and the luminosities for gamma-gamma \(\mathcal{L}_{\gamma\gamma}\), electron-electron \(\mathcal{L}_{ee}\), and electron-gamma \(\mathcal{L}_{e\gamma}\). The value of \( N_\gamma = 1.478 \times 10^{12}/\mathrm{s} \) from GBET agrees excellently with both the CAIN result (\( 1.465 \times 10^{12}/\mathrm{s} \)) and the theoretical prediction (\( 1.534 \times 10^{12}/\mathrm{s} \)). The GBET results for the luminosities $\mathcal{L}_{\gamma\gamma}$, $\mathcal{L}_{ee}$, and $\mathcal{L}_{e\gamma}$ agree with the CAIN results within 2\%, with relative differences of 1.76\%, 0.12\%, and 0.78\% respectively. Fig.~\ref{Fig11} presents a comparison of the energy spectra from GBET and CAIN: (a)~Compton photon energy spectra and (b)~spectra of all electrons after scattering. The excellent agreement across all compared parameters validates the effectiveness and accuracy of the GBET code.
\renewcommand{\arraystretch}{1.25}
\begin{table}[htbp!]
	\centering
	\caption{Comparison of simulation results between GBET and CAIN} \label{tab3}
	\begin{tabularx}{\textwidth}{
			>{\raggedright\arraybackslash\hsize=0.9\hsize}X
			>{\centering\arraybackslash\hsize=0.9\hsize}X
			>{\centering\arraybackslash\hsize=0.9\hsize}X
			>{\centering\arraybackslash\hsize=0.9\hsize}X
		}
		\toprule
		\textbf{Parameter} & \textbf{GBET} & \textbf{CAIN} & \textbf{Rel. Diff. (\%)} \\
		\midrule
		\( N_\gamma \) (s\(^{-1}\)) & \(1.4776 \times 10^{12}\) & \(1.4648 \times 10^{12}\) & 0.87 \\
		\( \mathcal{L}_{\gamma\gamma} \) (cm\(^{-2}\)s\(^{-1}\)) & \(5.1045 \times 10^{28}\) & \(5.1943 \times 10^{28}\) & 1.76 \\
		\( \mathcal{L}_{ee} \) (cm\(^{-2}\)s\(^{-1}\)) & \(1.1829 \times 10^{28}\) & \(1.1815 \times 10^{28}\) & 0.12 \\
		\( \mathcal{L}_{e\gamma} \) (cm\(^{-2}\)s\(^{-1}\)) & \(2.3810 \times 10^{28}\) & \(2.3625 \times 10^{28}\) & 0.78 \\
		\bottomrule
	\end{tabularx}
\end{table}
    
\section{Conclusion}\label{sec5}

In this work, a MC code, GBET, has been developed for simulating two sequential inverse Compton scattering processes in MeV Gamma-Gamma Collider. GBET overcomes the inherent information loss in conventional luminosity-spectrum-based chained simulations, preserving full particle-level information and achieving significantly higher physical fidelity. This improvement is important for  distinguishing signals from background processes. Based on the existing Gamma-Gamma Collider parameters, the luminosity $\mathcal{L}_{\gamma\gamma}$ is $5.10 \times 10^{28}\;\mathrm{cm^{-2}s^{-1}}$, which exceeds both the $e^{-}e^{-}$ luminosity of $1.18 \times 10^{28}\;\mathrm{cm^{-2}s^{-1}}$ and the $e^-\gamma$ luminosity of $2.38 \times 10^{28}\;\mathrm{cm^{-2}s^{-1}}$. The background particles caused by the first inverse Compton scattering are photons with energy less than 18 eV and a rate of \( \textnormal{6.24} \times \textnormal{10}^{\textnormal{7}} \)/s. The background particles produced by the second inverse Compton scattering have energies below 11 MeV, including electrons and gamma photons, with rates of 51.99/s and 0.99/s respectively. In addition, the background electrons from Møller scattering have energies of approximately 200~MeV with a rate of 0.56/s. The background electron-positron pairs produced via the Breit-Wheeler process have energies ranging from 511~keV to 720~keV, with production rates of 1312.2/s and 1314.3/s, respectively. This study can provide a reference for beam alignment and detector optimization in experiments. This work does not take into account the nonlinear Compton scattering process or the polarization of the beam. Further studies will be carried out to address these issues.

\bmhead{Acknowledgements}

\begin{description}
	\item[\normalfont\bfseries Funding:] This work is supported by Guangdong Provincial Key Laboratory of Advanced Particle Detection Technology(2024B1212010005); Guangdong Provincial Key Laboratory of Gamma-Gamma Collider and Its Comprehensive Applications(2024KSYS001); National Key Program for S\&T Research and Development (2023YFA1607200); The China Postdoctoral Science Foundation (Certificate Number 2025M773390). The High-performance Computing Public
Platform (Shenzhen Campus) of Sun Yat-sen University.
\end{description}
\begin{description}
	\item[\normalfont\bfseries Data availability:] The data will be made available on reasonable request.
\end{description}

\begin{description}
	\item[\normalfont\bfseries Competing interests:] The authors declare no conflicts of interest.
\end{description}


\bibliography{sn-bibliography} 

\end{document}